\documentclass[showpacs,11pt,%showkeys,twocolumn,
preprint,preprintnumbers,nofootinbib,
groupedaddress,superscriptaddress,amsmath,amssymb]{revtex4}

\usepackage[dvipdfm]{graphicx}
\usepackage{color,citesort}
\usepackage{cancel,graphics,subfigure}

\graphicspath{{figs/}}

\newcommand{\BR}{{\mathcal{B}}}

\begin{document}
%%%%%%%%%%%%%%%%%%%%%%%%%%%%%%%%%%%%%%%%%%%%%%%%%%%%%%%%%%%%%%%%%%%%%%%%%%%%%%%
\title{Doubly resonant $WW$ plus jet signatures at the LHC}
\preprint{}
%%%%%%%%%%%%%%%%%%%%%%%%%%%%%%%%%%%%%%%%%%%%%%%%%%%%%%%%%%%%%%%%%%%%%%%%%%%%%%%
\author{Johan Alwall}
\email{jalwall@ntu.edu.tw}
\affiliation{Department of Physics and Center for Theoretical
 Sciences, National Taiwan University, Taipei 10617, Taiwan}
%%%%%%%%%%%%%%%%%%%%%%%%%%%%%%%%%%%%%%%%%%%%%%%%%%%%%%%%%%%%%%%%%%%%%%%%%%%%%%%
\author{Tsedenbaljir Enkhbat}
\email{enkhbat@phys.ntu.edu.tw}
\affiliation{Department of Physics and Center for Theoretical Sciences,
National Taiwan University, Taipei 10617, Taiwan}
%%%%%%%%%%%%%%%%%%%%%%%%%%%%%%%%%%%%%%%%%%%%%%%%%%%%%%%%%%%%%%%%%%%%%%%%%%%%%%%
\author{Wei-Shu Hou}
\email{wshou@phys.ntu.edu.tw}
\affiliation{Department of Physics,
National Taiwan University, Taipei 10617, Taiwan}
%%%%%%%%%%%%%%%%%%%%%%%%%%%%%%%%%%%%%%%%%%%%%%%%%%%%%%%%%%%%%%%%%%%%%%%%%%%%%%%
\author{Hiroshi Yokoya}
\email{hyokoya@hep1.phys.ntu.edu.tw}
\affiliation{Department of Physics and Center for Theoretical Sciences,
National Taiwan University, Taipei 10617, Taiwan}
\affiliation{National Center for Theoretical Sciences, National Taiwan
University, Taipei 10617, Taiwan}
%%%%%%%%%%%%%%%%%%%%%%%%%%%%%%%%%%%%%%%%%%%%%%%%%%%%%%%%%%%%%%%%%%%%%%%%%%%%%%%

%\date{\today}

%%%%%%%%%%%%%%%%%%%%%%%%%%%%%%%%%%%%%%%%%%%%%%%%%%%%%%%%%%%%%%%%%%%%%%%%%%%%%%%
\begin{abstract}
  We present search prospects and phenomenology of doubly resonant
  signals that come from the decay of a neutral weak-singlet color-octet 
  vector state $\omega_8$ into a lighter weak-triplet color-octet
  scalar $\pi_8$, which can arise in several theories beyond the
  Standard Model. Taking $m_{\omega_8}-m_{\pi_8}>m_W$, we 
  demonstrate an analysis of the signals $pp\to\omega_8\to \pi^\pm_8
  W^\mp (\pi^0_8Z) \to g W^\pm W^\mp (gZZ)$.  The present 8 TeV LHC run
  is found to have the potential to exclude or discover the signal for
  a range of masses and parameters. The preferred search channel has a
  boosted $W$-tagged jet forming a resonance with a second hard jet,
  in association with a lepton and missing energy.
\end{abstract}
%%%%%%%%%%%%%%%%%%%%%%%%%%%%%%%%%%%%%%%%%%%%%%%%%%%%%%%%%%%%%%%%%%%%%%%%%%%%%%%
\maketitle

%%%%%%%%%%%%%%%%%%%%%%%%%%%%%%%%%%%%%%%%%%%%%%%%%%%%%%%%%%%%%%%%%%%%%%%%%%%%%%%

%%%%%%%%%%%%%%%%%%%%%%%%%%%%%%%%%%%%%%%%%%%%%%%%%%%%%%%%%%%%%%%%%%%%%%%%%%%%%%%
\section{Introduction}\label{intro}

The LHC experiments are now collecting data at an unprecedented rate,
bringing in an exciting period for potential new discoveries.
As a hadron collider, it is particularly apt at probing for new colored
objects.
Such particles are predicted in a variety of theories beyond the
Standard Model (SM), and include:  massive
gluons~\cite{Frampton:1987dn, Bagger:1987fz} including KK
gluon~\cite{Davoudiasl:2000wi, Lillie:2007yh, Appelquist:2000nn},
diquarks~\cite{Hewett:1988xc},  leptoquarks~\cite{Hewett:1988xc,
Dorsner:2005fq}, excited quarks~\cite{Baur:1987ga, 
Baur:1989kv}, squarks and gluino~\cite{Martin:1997ns}, colored version
of technicolor particles~\cite{Chivukula:1995dt} and other composite
colored objects~\cite{Hill:1991at, Hill:1993hs, Han:2010rf, Wise,
Enkhbat:2011vp, Dobrescu:2007yp, Bai:2010mn, 
Schumann:2011ji, Dobrescu:2011px, Kats:2012ym}. 
Currently, many of these objects have been searched for by
the CMS and ATLAS collaborations at the LHC,
with lower bounds~\cite{CMSdijet, Atlasdijet, :2012tx,
Chatrchyan:2012jx, :2012mf, :2012pq} placed on their masses. 
Among composite colored objects, a neutral SU(2)$_L$ singlet color-octet
vector resonance, as predicted by many models~\cite{Frampton:1987dn,
Davoudiasl:2000wi, Lillie:2007yh, Hill:1991at, Hill:1993hs, Han:2010rf,
Wise, Enkhbat:2011vp}, is of particular interest,
since it can be produced singly.
These objects often cascade decay to other lower lying states,
leading to distinct collider signals.
With masses in the TeV range, they often lead to boosted $t$'s and
$W$'s in the final states.
The study of such cascade decays is important in its own right, since it
differs characteristically from studies for a single resonance decaying
to a pair of light SM particles.
Whatever beyond the SM physics we consider, it often contains not just
one but multiple particles with various different quantum numbers,
and may lead to a double resonant signal.

In the current study, we examine the phenomenology of an SU(2)$_L$
singlet color-octet vector particle accompanied by a lighter SU(2)$_L$
triplet color-octet scalar. 
The weak SU(2)$_L$ quantum number, as an isospin symmetry, often plays a
crucial role for classifying possible spectrum.
A prevailing feature among many dynamical electroweak symmetry breaking
models is that they have isotriplets as the lower lying states,
accompanied by heavier isosinglets.
For example technicolor models, whether SU(3)$_c$ colored versions or
not, have spectra with $\pi$ and $\rho$-like resonances as the lightest
states, followed by an $\omega$-like resonance.
This spectrum also naturally arises when considering states composed of
a pair of SU(2)$_L$ doublet color-triplet spin-half particles, such as
the scenario studied in Ref.~\cite{Enkhbat:2011vp}.

As a case study for a signal that has not been considered previously,
and as an excellent example of a signal that can naturally
appear when studying resonant structures with more than one colored
state, we consider $\omega_8\to\pi_8 W/Z$ followed by $\pi_8 \to g W/Z$,
assuming $m_{\omega_8}-m_{\pi_8}>M_W$. While such spectrum has
appeared in several scenarios beyond the SM, phenomenological
studies thus far have been restricted to one resonance at a
time~\cite{Lillie:2007yh, Baur:1987ga, Chivukula:1995dt, Han:2010rf,
Bai:2010mn, Dobrescu:2011px}.
Thus, the final decay products we consider are $\omega_8\to(\pi^\pm_8
W^\mp,\,\pi^0_8Z)\to (g W^\pm W^\mp,\, gZZ)$.
With the $\omega_8$ mass in the TeV range and a $\pi_8$ mass of
several hundred GeV, this decay chain leads to a spectacular signal with
a very energetic hadronic jet and at least one highly boosted $W$ or $Z$
boson.
To take full advantage of the presence of such boosted objects in our
signal, we employ techniques for tagging boosted objects using jet
substructure~\cite{Butterworth:2008iy,Cui:2010km,Altheimer:2012mn}.
This also allows for an excellent reconstruction of the invariant mass
of such objects, once they have been discovered.

The paper is organized as follows. In Section~II, we give
the details about the signature, and discuss the
methods we use for tagging boosted weak bosons that decay hadronically,
and for determining exclusion and discovery limits. Details of the
numerical simulation and results are given in Section~III, and in
Section~IV, we give conclusions and discussion.

%%%%%%%%%%%%%%%%%%%%%%%%%%%%%%%%%%%%%%%%%%%%%%%%%%%%%%%%%%%%%%%%%%%%%%%%%%%%%%%

%%%%%%%%%%%%%%%%%%%%%%%%%%%%%%%%%%%%%%%%%%%%%%%%%%%%%%%%%%%%%%%%%%%%%%%%%
\section{Model Setup and Collider Signatures }

The scenario we consider has an SU(2)$_L$ singlet color-octet vector boson
$\omega_8$ that decays to an SU(2)$_L$ triplet color-octet scalar boson $\pi_8$
plus an SU(2)$_L$ gauge boson. The $\pi_8$ boson subsequently decays to
a gluon and another weak boson, leading to a doubly resonant signal
$W^\pm(W^\mp+\mathrm{jet})$ or $Z(Z+\mathrm{jet})$. While additional
color-singlet states might be present in the spectrum depending on
the details of the model, we disregard them in this study, since the
production cross section of such color-singlet particles typically would
be considerably smaller than that of the states discussed here.

Note that single production of $\pi_8$ from gluon fusion is forbidden by its isotriplet nature.
Although in principle $q\bar q$ initial states can be allowed, they, and the corresponding dijet decays, are usually suppressed in dynamical models. 
This is because the couplings to the light generations are 
usually proportional to the Yukawa couplings
since the color-singlet counter parts are the Goldstone bosons.
This still leaves the possibility of decay to $t\bar t$.  However, inducing a large enough top  
Yukawa coupling in these models is usually problematic and often invokes some extra mechanism.
We here ignore this possibility and restrict our phenomenological study to $\pi_8\to Wg$, which should always be present based on quantum numbers alone. This decay also naturally appears in composite models such as the one studied in Ref.~\cite{Enkhbat:2011vp}.

\begin{figure}[tb]
 \centering
 \includegraphics[height=4cm]{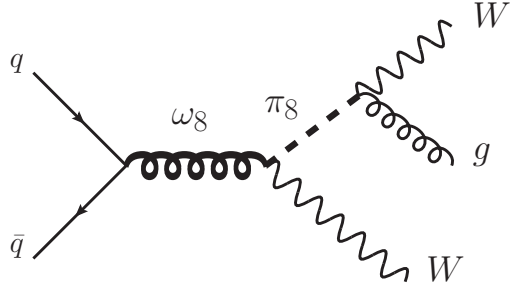}
 \caption{Feynman diagram of the $q\bar q\to\omega_8 \to \pi_8 W\to gWW$
 process.}\label{Fig:Feyn}
\end{figure}

The $\omega_8$ on the other hand is an SU(2)$_L$ singlet vector boson, and
can be produced singly in $q\bar q$ fusion (gluon fusion production is
forbidden by Yang's theorem~\cite{Yang:1950rg}). Therefore at hadron colliders, the
leading production mechanism of new states in this scenario, for sufficiently heavy
$\pi_8$, is single $\omega_8$ production via quark anti-quark
annihilation, $q \bar q\to \omega_8$.
The corresponding Feynman diagram is drawn in Fig.~\ref{Fig:Feyn}.

%--------------------------------------------------------------------------
%--------------------------------------------------------------------------

Besides the decay $\omega_8\to \pi_8 W/Z$, $\omega_8$ can also decay
to $q\bar q$ ($t\bar t$), as required by the unitarity relation. 
Thus the interactions that give the leading decays are expressed by the
following effective operators: 
\begin{eqnarray}
{\cal L}_{\omega_8\mbox{--decay} }&=& \xi \frac{g_s}{\sqrt{2}}\bar q_i \gamma_\mu (T^a)_{ij} q_j V_{a}^{\mu}+\frac{g_W}{\sqrt{m_{S}m_{V}}}\left[f_{v}\epsilon^{\mu\nu\rho\sigma}\left(\partial_\sigma S^\pm_a\partial_\rho W^\mp_\mu +\partial_\sigma S^0_a\partial_\rho Z_\mu\right)V_{a\nu}\right.\nonumber\\
&+&\left. S^\pm_a\left(f_{a_1}m_{S}m_{V}V_a^\mu+f_{a_2}\partial^\mu V_{a\nu} \partial^\nu +f_{a_3}V_{a\nu}\partial^\mu \partial^\nu   \right)W^\mp_\mu\right.\nonumber\\
&+&\left. S^0_a\left(f_{a_1}m_{S}m_{V}V_a^\mu+f_{a_2}\partial^\mu V_{a\nu} \partial^\nu +f_{a_3}V_{a\nu}\partial^\mu \partial^\nu   \right)Z_\mu\right].\label{effLag}
\end{eqnarray}
where
$\xi \equiv f_{V}/m_{V}$ is the decay constant (in units of $m_V$) for
the color-octet vector $V$,
$f_v$ and $f_{a_i}$ are form factors for the effective operators
connecting $V$ with the SU(2)$_L$ triplet color-octet scalar $S$. 
Note that SU(2)$_L$ indices have been suppressed.
The decay constant parameter $\xi$ fixes the production cross-section
and the dijet rate. 
The form factors $f_{v,\,a_i}$ control the transition decay of
$\omega_8$ to $\pi_8$'s. 
In the analysis given in the next Section, we assume that 
$f_{v,\,a_i}$ are large enough compared to $\xi$ for the decay into
$\pi_8 W/Z$ to dominate over the dijet decay.
Note that isospin invariance fixes the ratio of $\BR(\omega_8\to\pi_8^0
Z)/\BR(\omega_8\to\pi_8^\pm W^\mp)$ to be $1:2$,
where we ignore the $W$ and $Z$ mass difference.

If $\xi$ is large enough relative to $f_{v,\,a_i}$ to get a
significant branching ratio into $q\bar q$, the value of $\xi$ can be
constrained by the dijet resonance searches.
Stringent limits on the high-mass dijet resonance production
cross-section have been obtained by CMS with 5~fb$^{-1}$
data~\cite{CMSdijet}.
We convert this limit into an upper limit for $\xi$ by using the
first term of Eq.~(\ref{effLag}) to calculate the single $\omega_8$
cross-section at the LHC.
The partonic cross-section for this process can be written
as~\cite{Enkhbat:2011vp} 
\begin{align}
 \hat\sigma_{q\bar{q}\to\omega^{}_8}(\hat{s}) =
 \frac{32\pi^3\alpha_s^2}{9m_{\omega_8}^2} \xi^2
 \delta{(1-m_{\omega_8}^2/\hat{s})}.
\end{align}
Assuming the 100\% branching ratio of $\omega_8$ into dijet, $\xi$ is
constrained to be less than $\sim 0.1$ ($\sim 0.2$) for $m_{\omega_8}=
1000$ (2000)~GeV.
However, if the branching ratio into dijet is assumed to be small,
the limit is considerably weakened.

%--------------------------------------------------------------------------
\begin{figure}[tb]
 \centering
 \includegraphics[height=5.0cm]{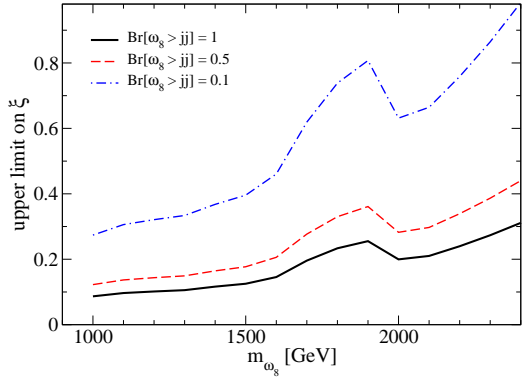}
 \caption{Upper limit on $\xi = f_{\omega_8}/m_{\omega_8}$ from the
 CMS dijet resonance search with ${\cal L}=5$~fb$^{-1}$~\cite{CMSdijet}.
 Several lines correspond to different assumptions of the branching
 ratio of $\omega_8$ into dijet. }\label{Fig:UL_xi}
\end{figure}
%--------------------------------------------------------------------------

In Fig.~\ref{Fig:UL_xi}, we plot the upper limit on $\xi$ obtained from the
CMS dijet data with 5~fb$^{-1}$~\cite{CMSdijet} by assuming several
values of the branching ratio of $\omega_8$ into dijet.
To calculate the hadronic cross-section at the LHC, we have used the
CTEQ6L1 parton distribution functions~\cite{CTEQ6} with the factorizing
scale $\mu_F=m_{\omega_8}$.
It is clear that, so long that the dijet does not dominate $\omega_8$ decay,
the bound on $\xi$ is very accommodating.
This illustrates the importance for resonances searches beyond the simple
dijet decays. 

In our analysis, we assume that the $\pi_8$ has a mass above 600 GeV,
for several reasons: If the $\pi_8$ is light, it would have a large
$gg \to \pi_8\bar\pi_8$ pair production cross section at the LHC, leading
to a different phenomenology~\cite{Dobrescu:2007yp,Schumann:2011ji} than
we are considering here.
Furthermore, we want to consider the region where the $W/Z$ from
the $\pi_8$ decay is sufficiently boosted for $W$-jet tagging to be
effective.

\subsection{Collider signatures}

The $W$- ($Z$-) boson from $\pi_8$ decay is expected to be very
energetic, since the energy is half the resonance mass in the
$\pi_8$ rest-frame.  On the other hand, the energy of the $W$-
($Z$-) boson of the $\omega_8$ decay is characterized by the mass
difference between the two states. For composite particles, we can
expect the mass splitting to be considerably smaller than the mass of
each of the states.
The $WWg$ ($ZZg$) signal in this case has one hard vector boson
with energy of order half the resonance mass, and one soft vector boson
with energy characterized by the mass difference between the resonances.
Although in the next Section, we extend our analysis to allow for a
free mass ratio between the new states, we will in the following refer
to the $W$ ($Z$) from the $\pi_8$ decay as the ``hard" vector boson,
and that from the $\omega_8$ decay as the ``soft" vector boson.

We first consider the case when $\omega_8\to\pi_8^\pm W^\mp$.
When both $W$-bosons decay hadronically, it is difficult to
distinguish the signal from the large QCD multijet background.  The
semi-leptonic decay is more promising, due to the presence of an
isolated lepton and missing energy, which significantly
reduces the background. It has two cases:
The hard $W$ decays hadronically while the soft $W$ decays
leptonically, where we consider only decays to $e$ and $\mu$, and vice versa.
The former case has an isolated charged lepton plus sizable missing
transverse momentum with three energetic jets, two of which are highly
collimated due to the large boost of the $W$. In this case, it is
useful to analyze the hadronically decaying $W$ as a single ``fat'' jet
with a substructure of two ``subjets''. Some methods for doing this will
be discussed in Sec.~\ref{sec:Wjets} below.
The latter case is characterized by a highly energetic isolated
charged lepton plus large missing transverse momentum, one very
energetic jet and two more jets which have an invariant-mass
close to $m_W$.  Due to the large boost, the directions of the charged
lepton and the missing transverse momentum in the transverse plane are
close to each other.

Finally, dileptonic decay $WWg\to (\ell\nu)(\ell\nu)j$ would be seen
as two charged leptons plus large missing transverse momentum and one
very energetic jet. The dileptonic decay ratio into $e$ and $\mu$ is
only about 5\%, so too small to be interesting in a discovery study.
Furthermore, the reconstruction of the final state is more difficult
in the dilepton case.

We next consider the decay of $\omega_8$ into a neutral $\pi_8^0$ plus
a $Z$ boson, giving a final state of $ZZg$.
The semileptonic cases are $ZZg\to (\ell^+\ell^-)(jj)j$ and
$(jj)(\ell^+\ell^-)j$, where the first (second) parenthesis denotes
the decay of the soft (hard) $Z$-boson.
In the case where the hard $Z$-boson decays leptonically, the leptons
are highly collimated, therefore only the muonic decay may possibly be
measured with some precision.  In the case of hadronic
decay of the hard $Z$ boson, the dijet will again be collimated, and
the methods discussed in Sec.~\ref{sec:Wjets} below can be used.
In order to reduce the large backgrounds from dileptonic $t\bar t$
decay, the analysis in Sec.~\ref{sec:simulation} requires the dilepton
invariant mass to be close to $m_Z$, and the missing energy to be
small.

In Sec.~\ref{sec:simulation}, we will perform a detailed analysis of
these collider signatures.

\subsection{$W$-jet tagging}
\label{sec:Wjets}

We explain here the $W$-tagging method used in our study.
We follow the method laid out in Ref.~\cite{CMSJetSub}.
Similar detailed studies can be found e.g.\ in
Refs.~\cite{Cui:2010km,ATLASBoost}.
According to Ref.~\cite{CMSJetSub}, the $W$-tagging method is summarized
as follows:
\begin{enumerate}
 \item Find jets by the C/A algorithm with $R=0.8$; keep the clustering
       history and momenta of clusters to be merged at each step.
 \item Perform pruning~\cite{Ellis:2009su}: Rerun the clustering; at each step,
       check if the two clusters $a$ and $b$ satisfy the two conditions,
\begin{align}
 & z_{ab}\equiv\frac{{\rm min}(p_T^a,p_T^b)}{p_T^J} < z_{\rm cut}, \\
 & \Delta R_{ab} > D_{\rm cut} \equiv \alpha\cdot\frac{M_J}{p_T^{J}},
\end{align}
       where $\Delta
       R_{ab}=\sqrt{(\Delta\eta_{ab})^2+(\Delta\phi_{ab})^2}$ and $M_J$
       is the mass of the jet in the original clustering sequence.
       If so, throw away the softer cluster instead of merging.
       The default values~\cite{Ellis:2009su} $z_{\rm cut}=0.1$ and
       $\alpha=1$ are taken for the pruning parameters.

 \item Mass drop tagging~\cite{Butterworth:2008iy}: Require the pruned
       jet mass to satisfy 60 GeV $<M_{\rm jet}<100$~GeV; the jet is tagged
       as a $W$ candidate if there exists a mass drop,
\begin{align}
 \frac{M_1}{M_{\rm jet}}<0.4,
\end{align}
       where $M_1$ is the mass of the hardest cluster in the last step
       of the pruning procedure.
\end{enumerate}

\subsection{The $\mathrm{CL}_s$ method}
\label{sec:CLs}

In the analysis described in the following section, we use the
$\mathrm{CL}_s$ method \cite{Read:2000ru} to determine the expected
cross section exclusion and discovery prospects at the 20 fb${}^{-1}$
early LHC run of 2012.  The $\mathrm{CL}_s$ method was developed for
Higgs searches by the LEP experiments, and it is the standard method
used for determining exclusion limits at the LHC. 
The fundamental quantity used to determine the
exclusion level is
\begin{equation}
  \label{eq:CLs}
  C\!L_s\equiv C\!L_{s+b}/C\!L_b
\end{equation}
where $C\!L_{s+b}=P_{s+b}(Q \le Q_{\rm obs})$ and $C\!L_b=P_{b}(Q \le
Q_{\rm obs})$ are the probabilities for some test-statistic to be less
than or equal to the observed value. This definition allows for a
consistent frequentist statistical treatment of experimental results
in the presence of backgrounds. An exclusion at confidence level $C\!L$
is reached when $1-C\!L_s \le C\!L$.

As test-statistic, we use the likelihood ratio
$Q=\mathcal{L}(\vec{X},s+b)/\mathcal{L}(\vec{X},b)$ for a given
experimental result $\vec{X}$. For a result presented as a binned
histogram with $N_{\rm chan}$ bins, this likelihood ratio can be expressed
as
\begin{equation}
  \label{eq:likelihood}
  Q=e^{-s_{\rm tot}}+\prod_{k=1}^{N_{\rm chan}}\left(1+\frac{s_i}{b_i}\right)^{n_i}
\end{equation}
Here, $s_{\rm tot}$ is the total signal rate for all channels, $s_i$ and
$b_i$ are the number of signal and background events in bin $i$, and
$n_i$ is the number of observed events in bin $i$.

The probability distributions $P_{s+b}(Q \le Q_{\rm obs})$ and $P_{b}(Q
\le Q_{\rm obs})$ are determined using test experiments starting from
signal and background histograms with associated statistical and
systematic uncertainties. To do this, and to determine the resulting
expected exclusion limits, we use the RooStat package
\cite{Moneta:2010pm}. Besides the statistical uncertainty from the
Monte Carlo statistics and expected event statistics, we also assume a
20\% systematic uncertainty on the SM backgrounds.

For a discovery analysis, the procedure is the same, except $C\!L_{s+b}$
is used instead of $C\!L_s$.

%%%%%%%%%%%%%%%%%%%%%%%%%%%%%%%%%%%%%%%%%%%%%%%%%%%%%%%%%%%%%%%%%%%%%%%%%%%%%%%

\section{Numerical Simulations and Results}
\label{sec:simulation}

For the numerical analysis of the signatures described in the previous
sections, we simulate both signal and the SM backgrounds
using MadGraph 5~\cite{Alwall:2011uj} interfaced with PYTHIA
6.4~\cite{Sjostrand:2006za} for hadronization and underlying event, and
Delphes 1.9.2~\cite{Ovyn:2009tx} for detector simulation.
The signal model is implemented directly in the UFO
format~\cite{Degrande:2011ua}.
Both signal and backgrounds are simulated using the jet matching techniques
described in Ref.~\cite{Alwall:2008qv}.
For the detector simulation, the CMS card included in the MadGraph
distribution is employed with the modification that the jet
reconstruction scheme used is anti-$k_T$~\cite{Cacciari:2008gp} and
the jet radius is set to 0.4.  
$W$-tagged jets are reconstructed by using
FastJet~\cite{Cacciari:2005hq} starting from the calorimetric clusters
in Delphes by the method described in the previous section, and 
anti-$k_T$ jets corresponding to such jets are removed from the hard jet list. 
As a crosscheck, both
simulations and analyses are done independently by two of us using two
independent analysis programs.

Since the different signatures have quite different properties, we will
go through them in sequence in the following.

\subsection{Hard $W$-tagged jet with soft leptonic $W$}
\label{sec:HardWjet}

The first, and most promising, decay mode that we analyze is
$\omega_8\to\pi_8^{\pm}(W^\mp)_\mathrm{leptonic}$ with $\pi_8^\pm\to
(W^\pm)_\mathrm{hadronic} + \mathrm{jet}$, where ``leptonic'' includes
decays to $e^\pm$ and $\mu^\pm$.
In this decay mode, we have a lepton and missing energy giving excellent
trigger and QCD multijet background separation.
The hadronic decay products of the $W^\pm$ from $\pi_8$ decay are
collimated and can be analyzed as a $W$ jet, as described in
section~\ref{sec:Wjets}.
Note that, in our analysis we do not make any stringent assumption
regarding the $\omega_8$-$\pi_8$ mass splitting, except that it is
larger than the $W$ mass.

The main SM backgrounds for this signature are $W^\pm$
associated with hard QCD radiation faking a $W$ jet, and
semi-leptonically decaying $t\bar t$, where the $W$ jet comes from
either a boosted hadronically decaying $W$ or is being faked by a hard
$b$ jet.
The irreducible background $W^\pm$ in association with a $W$ or $Z$
boson and jets is only a few \% of the main backgrounds.

\begin{figure}[tb]
 \centering
 \includegraphics[width=0.475\columnwidth]{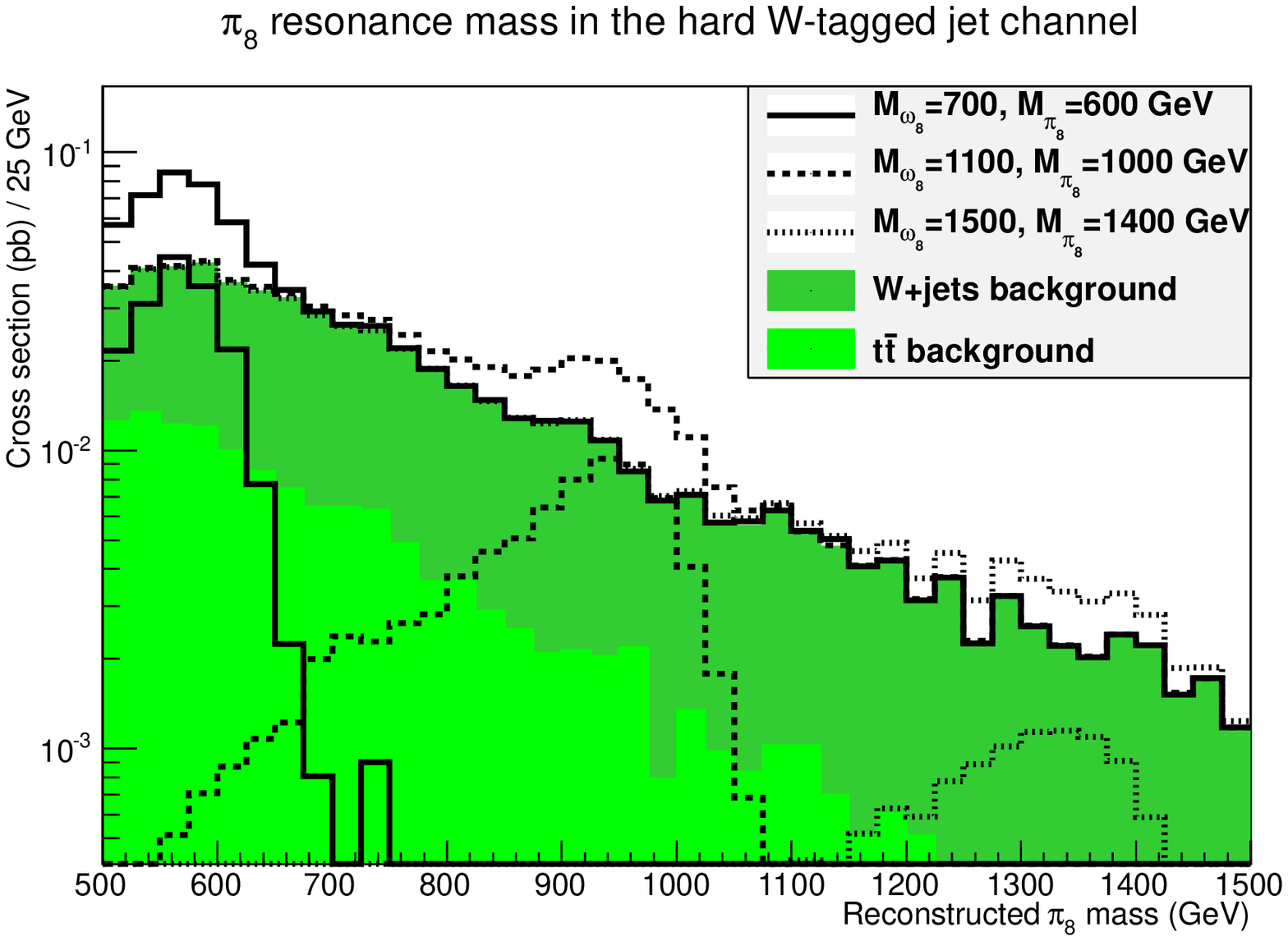}
 \includegraphics[width=0.475\columnwidth]{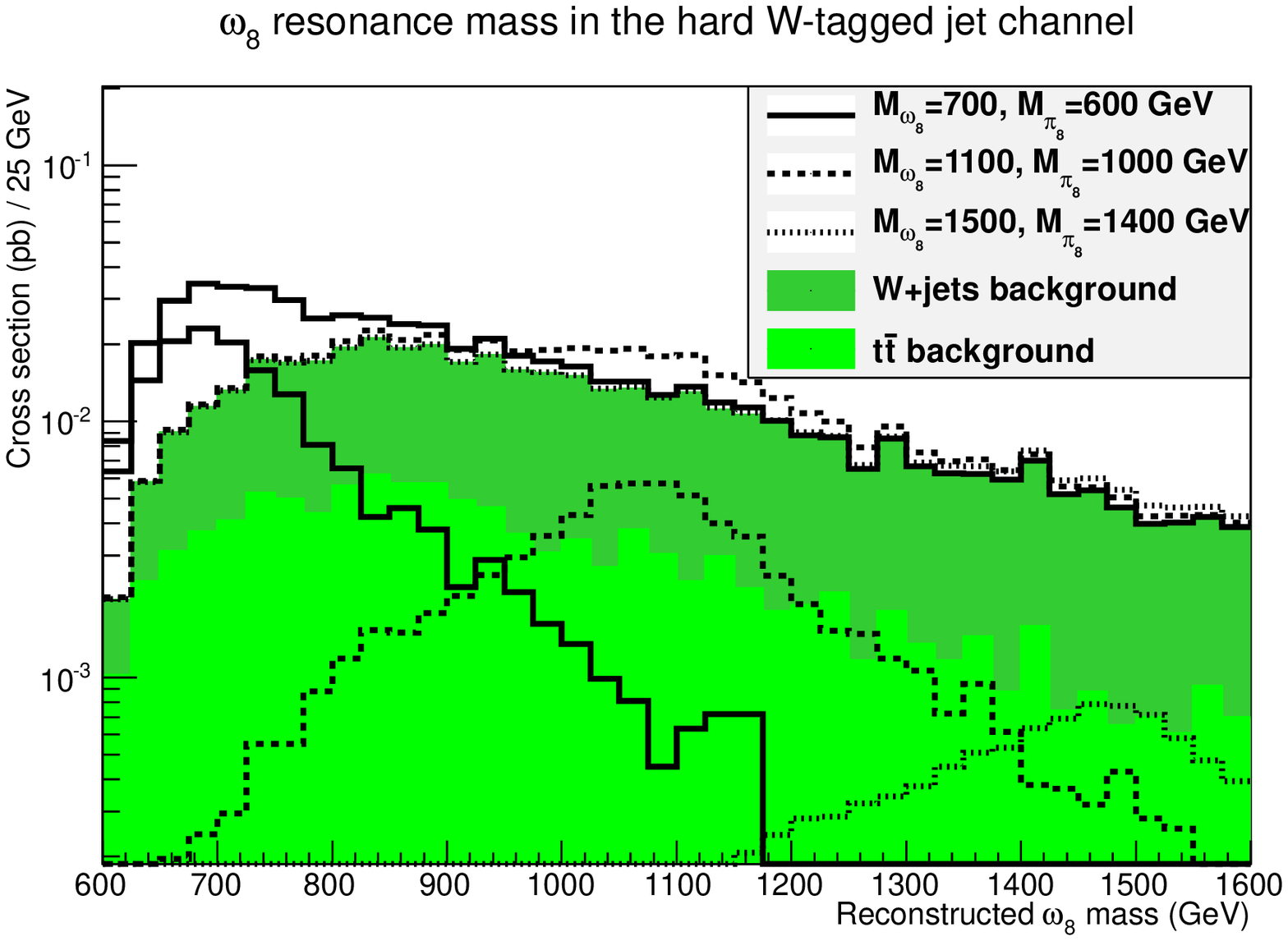}
 \caption{\label{fig:hadronic_masses} Invariant mass distributions for
   the reconstructed $\pi_8$ and $\omega_8$ for a few sample mass
   points in the hard $W$-tagged jet channel (see the text for
   details). The $\omega_8$ decay constant $\xi$ parameter is set to 0.2, in
   order to make the curves clearly distinguishable on top of the
   backgrounds.}
\end{figure}

We use the following set of cuts for our event selection:
\begin{itemize}
\item Exactly 1 isolated lepton ($e$ or $\mu$) with $p_\perp > 20$ GeV;
\item Missing $E_T > 20$ GeV;
\item A $W$-tagged jet $j_W^{}$ with $p_\perp > 200$ GeV;
\item A jet $j_1$ with $p_\perp > 200$ GeV;
\item Invariant mass (reconstructed $\pi_8$ mass) $M(j_W^{},j_1) > $ 500 GeV.
\end{itemize}
Varying the jet $p_\perp$ cuts only has a minimal effect on the
resulting exclusion curves.

The reconstructed $\pi_8$ and $\omega_8$ masses are shown for a few
different $(m_{\omega_8},m_{\pi_8})$ mass points in
Fig.~\ref{fig:hadronic_masses}, together with the main backgrounds.
We use $\xi=0.2$ in the figure, to
make the curves easily visible on top of the background. Note that the cross section scales as $\xi^2$,
since we assume 100\% branching fraction into the $\pi_8$ and a weak boson final state.
The width of the invariant mass distributions is mainly due to the
relatively large energy scale uncertainties for the reconstructed $W$ jet,
and the jet energy scale uncertainty for the gluon jet.
To reconstruct the $\omega_8$ mass, we first need to perform leptonic
$W$ reconstruction.
In the reconstruction of the leptonic $W$, we pick the solution with the
smallest neutrino $z$ momentum.
We furthermore allow a reconstructed $W$ mass up to 100 GeV, and if
necessary adjust the magnitude of the missing $E_T$ to bring the $W$
mass back to the nominal value.

In the exclusion and discovery analysis, we use the reconstructed $\pi_8$ mass,
since the peak resolution of the $\pi_8$ is considerably
better than the $\omega_8$ case. Note however that the double-resonant assumption is
crucial for the background suppression as described above.

\begin{figure}[tb]
 \centering
 \includegraphics[width=0.55\columnwidth]{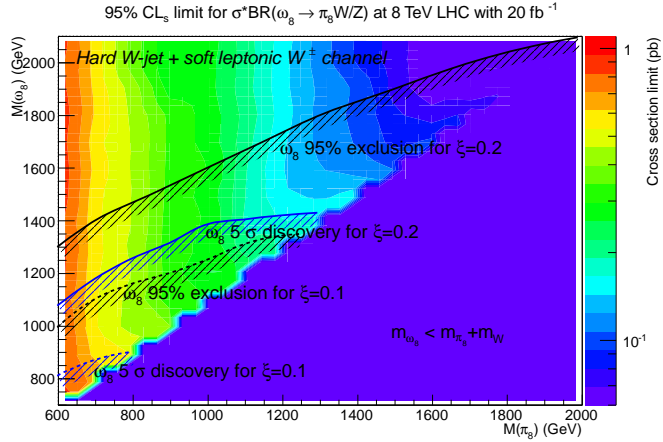}
 \caption{\label{fig:hadronic_exclusion} Cross section exclusion
 limits in pb in the $(\omega_8, \pi_8)$ mass plane in the hard
 $W$-tagged jet channel, for 20 fb${}^{-1}$ integrated luminosity at
 the 8 TeV LHC.
 Also the exclusion regions for $\omega_8$ with $\xi$ set to 0.1 and 0.2
 are indicated.}
\end{figure}

In Fig.~\ref{fig:hadronic_exclusion} we give the 95\% C.L. cross section
exclusion contours for 20 fb${}^{-1}$ at the 8 TeV LHC, using the
$\mathrm{CL}_s$ method as discussed in Sec.~\ref{sec:CLs}.
We also show both the exclusion as well as $5\sigma$
discovery limits for $\omega_8$ production for two different values
of $\xi$, 0.1 and 0.2, assuming
$\BR(\omega_8\to\pi_8^{\pm}W^{\mp})=2/3$ and
$\BR(\omega_8\to\pi_8^0Z)=1/3$, and $\pi_8^\pm \to W+$
jet decay 100\% of the time.
For $\omega_8$ and $\pi_8$ mass difference close to $m_W$,
the LHC will be able to exclude
$\omega_8$ production up to 2100 GeV (1350 GeV) for $\xi=0.2$ (0.1)
with 20 fb${}^{-1}$ integrated luminosity, and make an $\omega_8$
discovery for masses up to 1400 GeV (900 GeV) for $\xi=0.2$ (0.1).

\begin{figure}[tb]
 \centering
 \includegraphics[width=0.475\columnwidth]{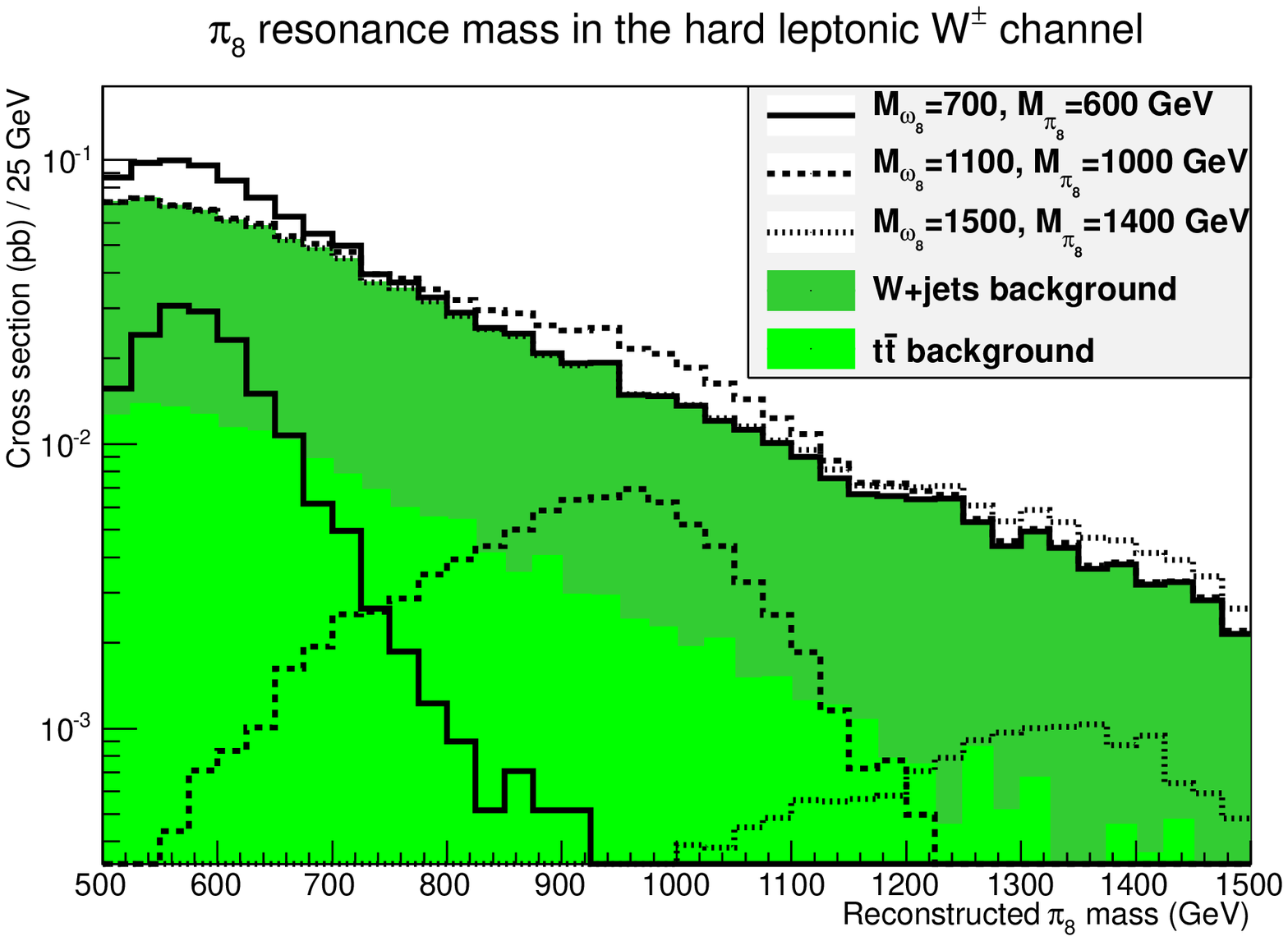}
 \includegraphics[width=0.475\columnwidth]{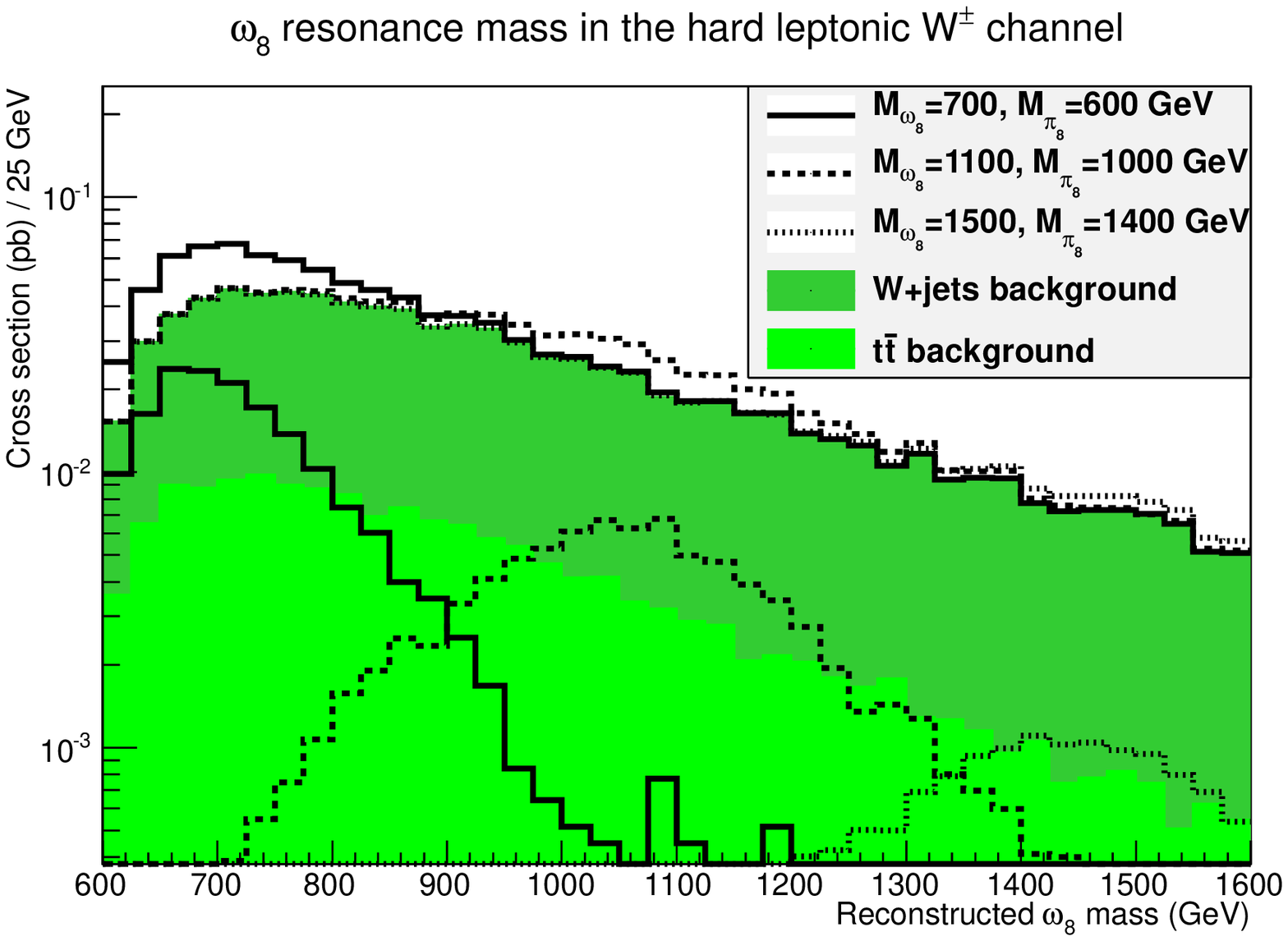}
 \caption{\label{fig:leptonic_masses} Invariant mass distributions for
 the reconstructed $\pi_8$ and $\omega_8$ for a few sample mass
 points in the hard leptonic $W$ channel (see the text for
 details).
 The $\omega_8$ decay constant parameter $\xi$ is set to 0.2, in
 order to make the curves clearly distinguishable on top of the
 backgrounds.}
\end{figure}

\subsection{Hard leptonic $W$ and soft hadronic $W$}
\label{sec:HardLepW}

The next decay mode we consider is
$\omega_8\to\pi_8^{\pm}(W^\mp)_\mathrm{hadronic}$ with $\pi_8^\pm\to
(W^\pm)_\mathrm{leptonic} + \mathrm{jet}$, where ``leptonic'' again
includes decays to $e^\pm$ and $\mu^\pm$ only.  While this decay mode
has exactly the same branching ratio suppression as the previously
discussed ``hard $W$ jet mode'', it lacks the distinguishing $W$ jet
signature.  On the other hand, it requires two jets in the event to
reconstruct to a hadronic $W$, which still gives significant reduction
of the main $W$ + jets background.  As in the previous case, the main
backgrounds are $W$ + jets and semileptonically decaying $t\bar t$.
The cuts used in this case are:
\begin{itemize}
\item Exactly 1 isolated lepton with $p_\perp>50$ GeV;
\item Missing $E_T > 50$ GeV;
\item A reconstructed leptonic $W_\mathrm{lep}$ with $p_\perp > 200$ GeV;
\item A jet $j_1$ with $p_\perp > 200$ GeV;
\item Invariant mass (reconstructed $\pi_8$ mass) $M(W_\mathrm{lep},j_1)
      > $ 500 GeV;
\item A hadronic $W$ reconstructed from two non-leading jets with
      $p_\perp > 20$ GeV, with the invariant mass between 50 and 110 GeV.
\end{itemize}

Details about leptonic $W$ reconstruction are given in
Sec.~\ref{sec:HardWjet}. 
If multiple hadronic $W$ candidates are found within the required
invariant mass window, the candidate with mass closest to the $W$ mass
is selected.
The energy of the hadronic $W$ is rescaled to get the correct $W$
mass, keeping $\eta$, $\phi$ and $p_\perp$ fixed, before reconstruction
of the $\omega_8$ from the reconstructed $\pi_8$ momentum and the hadronic $W$.

Fig.~\ref{fig:leptonic_masses} shows the reconstructed $\pi_8$ and
$\omega_8$ invariant mass distributions for a few different
$(m_{\omega_8},m_{\pi_8})$ mass points (using $\xi=0.2$), together
with the main backgrounds.  The invariant mass widths are slightly
larger than those in the ``hard $W$ jet'' case, due to the larger
energy scale uncertainty of the missing $E_T$, as well as the
combinatoric uncertainty for selecting among possible hadronic $W$
candidates.
We again use the reconstructed $\pi_8$ ``bump'' to determine the
exclusion and discovery limits.

\begin{figure}[tb]
 \centering
 \includegraphics[width=0.55\columnwidth]{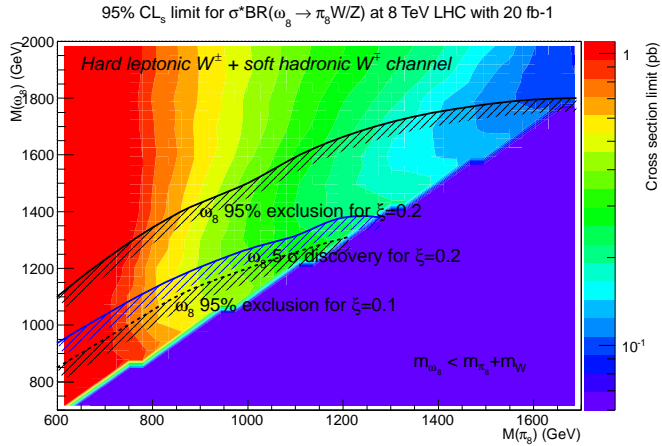}
 \caption{\label{fig:leptonic_exclusion} Cross section exclusion
 limits in pb in the $(\omega_8, \pi_8)$ mass plane in the hard
 leptonic $W$ channel, for 20 fb${}^{-1}$ integrated luminosity at
 the 8 TeV LHC.
 Also the exclusion regions for $\omega_8$ with $\xi$ set to 0.1 and 0.2
 are indicated.}
\end{figure}

Fig.~\ref{fig:leptonic_exclusion} shows the 95\% cross section
exclusion contours for 20 fb${}^{-1}$ at the 8 TeV LHC, as well as the
exclusion curves for $\omega_8$ production with two different values
of $\xi$, 0.1 and 0.2, assuming
$\BR(\omega_8\to\pi_8^{\pm}W^{\mp})=2/3$ and
$\BR(\omega_8\to\pi_8^0Z)=1/3$.  The exclusion reach for this
channel is, as expected, not as good as that for the hard $W$-tagged
jet channel described in the previous case (extending to
$m_{\omega_8}$ near 1300 GeV for $\xi=0.1$ and $m_{\omega_8}$ around
1800 GeV for $\xi=0.2$), but it can still be interesting as a
cross-check. For a $5\sigma$ discovery in this channel, the $\omega_8$
mass must be below 1380 GeV assuming $\xi=0.2$. As for $\xi=0.1$, no
discovery is possible in the studied mass region.

\subsection{Soft leptonic $Z$ with a hard $Z$ jet}

Since we assume the $\omega_8$ decay to be isospin invariant, we expect
$1/3$ of the events to feature $\omega_8 \to \pi_8^0 +Z$.
This channel might seem promising \emph{a priori}, thanks to the excellent invariant
mass reconstruction of a leptonically decaying $Z$ boson, as well as the
lower backgrounds. Here the $t\bar t$ background is negligible if a lepton
pair is required to be close to the $Z$ mass and an upper limit on
missing $E_T$ is imposed.
However, the smaller cross section for $Z$+jets production compared to
$W$+jets production for the background is balanced by several factors.
These include a considerably smaller, roughly a factor three, branching ratio into $e$ and $\mu$,
the factor two smaller $\omega_8$ decay rate, together with slightly worse cut
efficiencies (since, e.g., two isolated leptons are required).
Due to these factors, the $Z$ decay mode overall has a three times worse cross section exclusion than the $W$ mode, and will therefore not be relevant for an exclusion or discovery study.
But it would certainly be important in a later stage, after discovery of a
signal, to establish the isospin symmetry of the $\omega_8$ decays.

For completeness, here we list the cuts used in our analysis:
\begin{itemize}
\item Exactly 2 isolated same-flavor opposite-sign leptons ($e$ or
      $\mu$) with $p_\perp > 20$ GeV and an invariant mass
      $|M(l^+,l^-)-m_Z| < 15$ GeV;
\item Missing $E_T < 50$ GeV;
\item A $W$-tagged jet $j_Z^{}$ (with mass range centered around
      $m_Z$) with $p_\perp > 200$ GeV;
\item A jet $j_1$ with $p_\perp > 200$ GeV;
\item Invariant mass (reconstructed $\pi_8$ mass) $M(j_Z^{},j_1) > $ 500 GeV.
\end{itemize}

%%%%%%%%%%%%%%%%%%%%%%%%%%%%%%%%%%%%%%%%%%%%%%%%%%%%%%%%%%%%%%%%%%%%%%%%%%%%%%%
\section{Conclusions and Discussions}

In this article, we have studied search prospects and
the phenomenology of doubly resonant signals which result
from the decay of a neutral SU(2)$_L$ singlet color-octet vector state 
$\omega_8$ into a lighter SU(2)$_L$ triplet color-octet scalar $\pi_8$. 
While these type of resonances have been subject of
extensive collider studies, they have been mostly studied individually.

It should be noted that such color-octet resonances often
come together in various SU(2)$_L$ representations,
which give rise to doubly resonant signals.
Assuming $m_{\omega_8} > m_{\pi_8}+m_W$,
we have provided the search prospects and
phenomenology of $\omega \to \pi_8^\pm W^\mp
(\pi_8^0 Z)\to gW^\pm W^\mp (g ZZ)$ signature.
We find that the signal can be efficiently excluded, and even
discovered, for a large range of masses and parameter choices,
already for the present 8 TeV run of the LHC. 
The most effective search channel is for $\pi_8$ decaying into 
a boosted $W$ boson that decays hadronically, plus a hard gluon jet, while 
the $W$ from the $\omega_8$ decay goes into a lepton and missing
transverse energy. 
This channel gives at the same time excellent background 
suppression and superior resolution in the mass peak reconstruction.
Furthermore, for this channel, the dominant SM background is $W$ +
jets production, where a hard jet is falsely tagged as a $W$ jet. For
this channel, it might be possible to further improve the signal
significance by using more complicated multivariable $W$ jet tagging
methods, as described in Ref.~\cite{Cui:2010km}.

Our analysis demonstrates how a sizable branching fraction for
$\omega_8\to\pi_8W$ (rather than dominantly decaying into $q\bar q$) allows
access to two resonances in one stroke.  This is clearly
advantageous compared to the case where $\pi_8$ has to be produced in
pairs in strong interaction, making it inaccessible at the early run
at 8~TeV, in particular when the resonances have masses around or
above TeV. 
\\

\vspace{0.35cm}
%%%%%%%%%%%%%%%%%%%%%%%%%%%%%%%%%%%%%%%%%%%%%%%%%%%%%%%%%%%%%%%%%%%%%%%%%%%%%%%
\noindent\textbf{Acknowledgments.}
The authors would like to thank Kai-Feng Chen for useful discussions and 
patient help with the code for the $\mathrm{CL}_s$ method.
H.Y.\ thanks KEK Theory Center for the hospitality during his stay which
was made possible by the exchange program between KEK Theory Center and
NCTS in Taiwan.
J.A.\ is supported by NTU Grant No.~10R1004022, T.E.\ and H.Y.\ are
supported in part by the National Science Council of Taiwan under Grant
No.~NSC 100-2119-M-002-061 and NSC 100-2119-M-002-001. 
W.-S.H.\ thanks the National Science Council for Academic Summit grant
NSC~100-2745-M-002-002-ASP. 
%%%%%%%%%%%%%%%%%%%%%%%%%%%%%%%%%%%%%%%%%%%%%%%%%%%%%%%%%%%%%%%%%%%%%%%%%%%%%%%

%%%%%%%%%%%%%%%%%%%%%%%%%%%%%%%%%%%%%%%%%%%%%%%%%%%%%%%%%%%%%%%%%%%%%%%%%%%%%%%

\end{document}